\documentclass[aps,prl,reprint,superscriptaddress,twocolumn,longbibliography]{revtex4-1}
\usepackage{amssymb}
\usepackage{amsmath}
\usepackage{graphicx}
\usepackage{epstopdf}
\usepackage{times}
\usepackage[colorlinks, linkcolor=blue, urlcolor=blue, anchorcolor=blue, citecolor=blue]{hyperref}

\begin{document}
\title{Quantum-memory-enhanced preparation of nonlocal graph states}
\author{Sheng Zhang}\thanks{These authors contributed equally to this work.}
\author{Yu-Kai Wu}\thanks{These authors contributed equally to this work.}
\author{Chang Li} \altaffiliation{Present address:  ISIS (UMR 7006), University of Strasbourg and CNRS, 67000 Strasbourg, France}
\author{Nan Jiang}\altaffiliation{Present address: Department of Physics, Beijing Normal University, Beijing 100875, China}
\author{Yun-Fei Pu}
\author{Lu-Ming Duan}\email{lmduan@tsinghua.edu.cn}
\affiliation{Center for Quantum Information, IIIS, Tsinghua University, Beijing 100084, PR China}

\begin{abstract}
Graph states are an important class of multipartite entangled states. Previous experimental generation of graph states and in particular the Greenberger-Horne-Zeilinger (GHZ) states in linear optics quantum information schemes is subjected to an exponential decay in efficiency versus the system size, which limits its large-scale applications in quantum networks. Here we demonstrate an efficient scheme to prepare graph states with only a polynomial overhead using long-lived atomic quantum memories. We generate atom-photon entangled states in two atomic ensembles asynchronously, retrieve the stored atomic excitations only when both sides succeed, and further project them into a four-photon GHZ state. We measure the fidelity of this GHZ state and further demonstrate its applications in the violation of Bell-type inequalities and in quantum cryptography. Our work demonstrates the prospect of efficient generation of multipartite entangled states in large-scale distributed systems with applications in quantum information processing and metrology.
\end{abstract}

\maketitle

\begin{figure*}
  \centering
  \includegraphics[width=14cm]{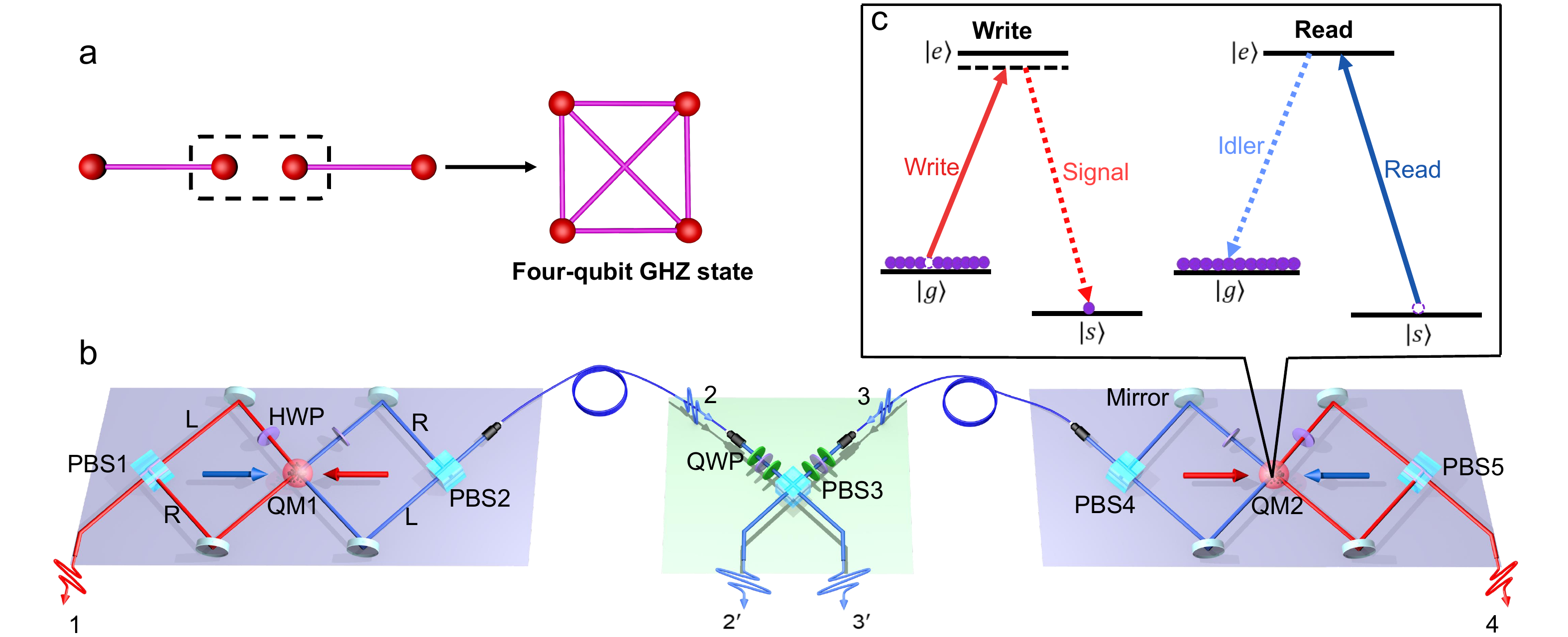}\\
  \caption{\textbf{Schematic for the memory-enhanced preparation of four-qubit GHZ states.}
  \textbf{a}, The scheme to create a four-qubit GHZ state. Two pairs of entangled qubits are generated asynchronously and are then projected into a four-qubit GHZ state. The two edge qubits not involved in the latter operations can be measured in advance depending on applications.
  \textbf{b}, The whole experimental setup consists of two symmetrically designed long-lived atomic quantum memories (QMs) and one intermediate interference station. Atom-photon entanglements are generated in the atomic ensembles  asynchronously using the DLCZ scheme by weak red-detuned write beams (red arrows). Upon registering two signal photons (1 and 4) successively, we apply strong resonant read beams (blue arrows) to retrieve the two atomic excitations into idler photons (2 and 3), and transmit them to the interference station through single mode fibers. A polarization beam splitter (PBS3) is used to project the two idler photons onto the subspace spanned by $|H\rangle|H\rangle$ and $|V\rangle|V\rangle$ if both output ports have photons (2$^\prime$ and 3$^\prime$). A sandwich structure of two quarter-wave plates (QWPs) and one half-wave plate (HWP) is used to compensate the polarization change of the photons during the fiber transmission.
  \textbf{c}, The energy level diagram for the write process and the read process. The relevant energy levels of the $^{87}$Rb atoms are $|g\rangle \equiv |5S_{1/2},F=2, m_F=0\rangle$, $|s\rangle \equiv |5S_{1/2},F=1, m_F=0\rangle$ and $|e\rangle \equiv |5P_{1/2},F=2\rangle$. The write beam is $20\,$MHz red detuned to the $|g\rangle \leftrightarrow |e\rangle$ transition while the read beam is resonant to the $|s\rangle \leftrightarrow |e\rangle$ transition.
  }
\end{figure*}

\emph{\textbf{Introduction.}}
As the available quantum devices scale up over the past few decades, multipartite entanglement has attracted intense research interest owing to its wide applications ranging from testing fundamental concepts such as quantum nonlocality \cite{GHZ1990,MABK1,*MABK2,*MABK3,rmp20steering} to the practical usage such as quantum computing \cite{shor1996fault,gottesman1999demonstrating,oneway}, quantum cryptography \cite{qss,pra03liar} and quantum metrology \cite{pra96optimal_frequency,sci04enhanced_measure,Toth12pra}. An archetypal class of the multipartite entangled states is the graph state \cite{Briegel04pra_graph}. It includes the cluster state with applications in measurement-based quantum computing \cite{oneway}, and the n-particle Greenberger-Horne-Zeilinger (GHZ) state \cite{GHZ1990} which is one of the maximally entangled states (see e.g. \cite{guhne2009entanglement}), shows largest violation in Bell-type inequalities \cite{MABK1,*MABK2,*MABK3,pra01Wolf}, and is used in various protocols in quantum information science \cite{shor1996fault,gottesman1999demonstrating,qss}. Graph states have previously been demonstrated in diversified physical systems. For example, the GHZ state has been generated in trapped ion systems for up to 24 qubits \cite{PRXQuantum.2.020343}, in superconducting circuits for up to 27 qubits \cite{Mooney_2021}, up to 20 qubits in Rydberg atom arrays \cite{lukin19rydberg}, up to 12 qubits in linear optical systems \cite{PhysRevLett.121.250505} (and up to 18 qubits using hybrid degrees of freedom of photons \cite{Pan18prl}), as well as in cavity QED systems \cite{Rauschenbeutel2024}, colored centers in diamonds \cite{nv08sci} and NMR systems \cite{pra10nmr}.

Despite the experimental progress, deterministic realizations of general graph states, and in particular the GHZ states, are currently restricted to local systems \cite{PRXQuantum.2.020343,Mooney_2021,lukin19rydberg}. For applications like distributed quantum computers and quantum networks \cite{kimble2008quantum,wehner2018quantum}, the linear optical system is preferred as it naturally supports long-distance transmission, which however suffers an exponential decay in the generation efficiency versus the system size due to the probabilistic entanglement operation \cite{PhysRevLett.121.250505,Pan18prl}. Incidentally, one-dimensional cluster states of photons have been generated with the help of a quantum emitter \cite{PhysRevLett.103.113602,doi:10.1126/science.aah4758}, but it is still challenging to generalize to higher dimensions for applications in measurement-based quantum computing.
To overcome this difficulty of scalable nonlocal graph states, theoretical proposals have been raised through the combination of atomic ensembles and linear optics \cite{duan02prl,duan06prl} inspired by the DLCZ protocol for quantum repeaters \cite{duan2001long}. The key idea is that, photons not involved in the later operations can be measured and postselected halfway by polarization beam splitters (PBSs) and single photon detectors (SPDs) \cite{duan06prl}; then large-scale graph states can be generated in a divide-and-conquer manner with the help of long-lived atomic quantum memories \cite{duan02prl}. Following this protocol, the exponential decay of the preparation efficiency versus the system size can be alleviated to only polynomial, which is crucially important for generating large-scale graph states using linear optics with relatively low entanglement generation and photon detection efficiencies. Similar ideas of memory assistance have also been explored in quantum-repeater-related studies for synchronizing single photons or preparing bipartite entanglement in various systems such as atomic ensembles \cite{felinto2006conditional,Chou07sci,pu2021experimental}, optical cavities \cite{Kaneda17optica,PhysRevLett.126.230506} and solid state spins \cite{lukin20nature}. However, due to the experimental difficulty such as low storage efficiency and fidelity, memory-enhanced generation of multipartite entanglement has not been realized yet.

In this work, we implement the memory-enhanced scheme to generate four-photon GHZ states. We load atomic ensembles in two optical traps to achieve a storage time of tens of milliseconds, produce atom-photon entanglement in each atomic memory asynchronously and then simultaneously convert the collective atomic excitations in the two memories into photons and project them to the desired four-photon GHZ state.
We show that the preparation efficiency of the GHZ state in the four-photon case is improved from a quadratic scaling to a linear one versus the generation rate of individual entangled pairs, which, when generalized to larger number of atomic memories and photons, leads to the desired polynomial scaling in efficiency rather than the exponential one.
We further measure the fidelity of this GHZ state and demonstrate its applications in the violation of MABK inequalities \cite{MABK1,*MABK2,*MABK3} and a quantum cryptography protocol of quantum secret sharing \cite{qss}. Our work realizes a prototype for the efficient preparation of a large-scale graph state, thus constitutes an important step towards its various applications in quantum information science and quantum metrology.

\emph{\textbf{Memory-enhanced generation of four-photon GHZ states.}}
Our experimental scheme is sketched in Fig.~1a. We generate two pairs of polarization-entangled photons and use one PBS and two SPDs to project them into a four-photon GHZ state \cite{duan06prl}. A crucial prerequisite for the efficient generation of multipartite GHZ states is to have long-lived quantum memories (QMs), such that the succeeded parts can be stored for long enough time until the other parts also succeed. In this experiment, we implement two quantum memories with $^{87}$Rb atomic ensembles loaded into one-dimensional optical lattices \cite{zhaoran08,yangsjlong,magic}. We have observed storage lifetime of tens of milliseconds \cite{supplementary}, which fully meets our requirement for the asynchronous preparation of entangled photon pairs.

As shown in Fig.~1b, we use the DLCZ scheme to create the atom-photon entanglement in the two atomic ensembles. We start from the QM1 with its atoms initially optically pumped to the ground state $|g\rangle$. A weak red-detuned write laser pulse can induce a spontaneous Raman transition $|g\rangle \rightarrow |s\rangle$ with a small probability $p$, which leads to the emission of a signal photon together with a collective spin wave excitation in the atomic ensemble, as shown in Fig.~1c. Here we collect the signal photon from two possible spatial modes $L$ and $R$ which locate symmetrically on the two sides of the write beam. We further use a half-wave plate (HWP) and a PBS to convert the path qubit into a polarization qubit (labelled as 1 in Fig.~1b) and register the signal photon with an SPD (we can measure it in any desired polarization basis depending on the applications). The effective atom-photon entangled state can be written as
\begin{equation}
|\Psi\rangle _{S-A}=\frac{1}{\sqrt{2}}(|H\rangle|L\rangle+e^{i\phi_S}|V\rangle|R\rangle), \label{EQ1}
\end{equation}
where $H/V$ denotes the horizontal/vertical polarization of the signal photon, $L/R$ the two spatial modes of the spin wave excitation, and $\phi_S$ the phase difference between the two signal paths before they are combined on the PBS1. Here we have ignored the large vacuum part of the state, which will be eliminated automatically by the postselection of photon detection; we have also dropped the higher-order excitations for simplicity owing to the low excitation probability.
This atom-photon entangled state can later be converted into photon-photon entanglement on demand by applying a strong resonant read pulse to retrieve the stored spin wave excitation into an idler photon (labelled as 2 in Fig.~1b). The resulting signal-idler photon entangled state can be described as
\begin{equation}
|\Psi\rangle_{12}=\frac{1}{\sqrt{2}}\left[|H\rangle |H\rangle +e^{i(\phi_S+\phi_I)}|V\rangle |V\rangle\right], \label{EQ2}
\end{equation}
where $\phi_I$ is the phase difference between the two idler paths before they are combined on the PBS2. In the experiment, the total phase $\phi_S+\phi_I$ is actively stabilized by a Mach-Zehnder interferometer. Thanks to the long-lifetime storage capacity, the atom-photon entanglement can be stored for milliseconds with a high entanglement fidelity above $90\%$ \cite{supplementary}.

Upon the successful entanglement generation in QM1 heralded by the detection of signal photon 1, we move on to create the atom-photon entanglement in QM2 in the same way. Once a signal photon 4 is detected, an atom-photon entangled state analogous to Eq.~(\ref{EQ1}) is prepared in QM2. After the asynchronous preparation of the two atom-photon entangled states, we apply two read beams to retrieve the atomic qubits into two idler photons simultaneously (labelled as 2 and 3) and direct them to the interference station with single mode fibers.

The core elements of the interference station are a PBS and two SPDs. Because the PBS transmits $H$ polarization and reflects $V$ polarization, the coincidence count between the two exits of PBS3 occurs only when both the idler photons $2^\prime$ and $3^\prime$ have the same polarization $H$ or $V$. In other words, the coincidence count projects the two idler photons into the subspace spanned by $|H\rangle|H\rangle$ and $|V\rangle|V\rangle$ \cite{duan06prl}. The product state $|\Psi\rangle_{12} \otimes |\Psi\rangle_{34}$ is thus projected into a four-photon GHZ state
\begin{equation}
 |\mathrm{GHZ}_4\rangle_{12^\prime 3^\prime 4}=\frac{1}{\sqrt{2}}(|HHHH\rangle +e^{i\phi} |VVVV\rangle),
\end{equation}
where the relative phase $\phi$ is compensated to be zero in the experiment \cite{supplementary}.

\begin{figure}
  \centering
  \includegraphics[width=7cm]{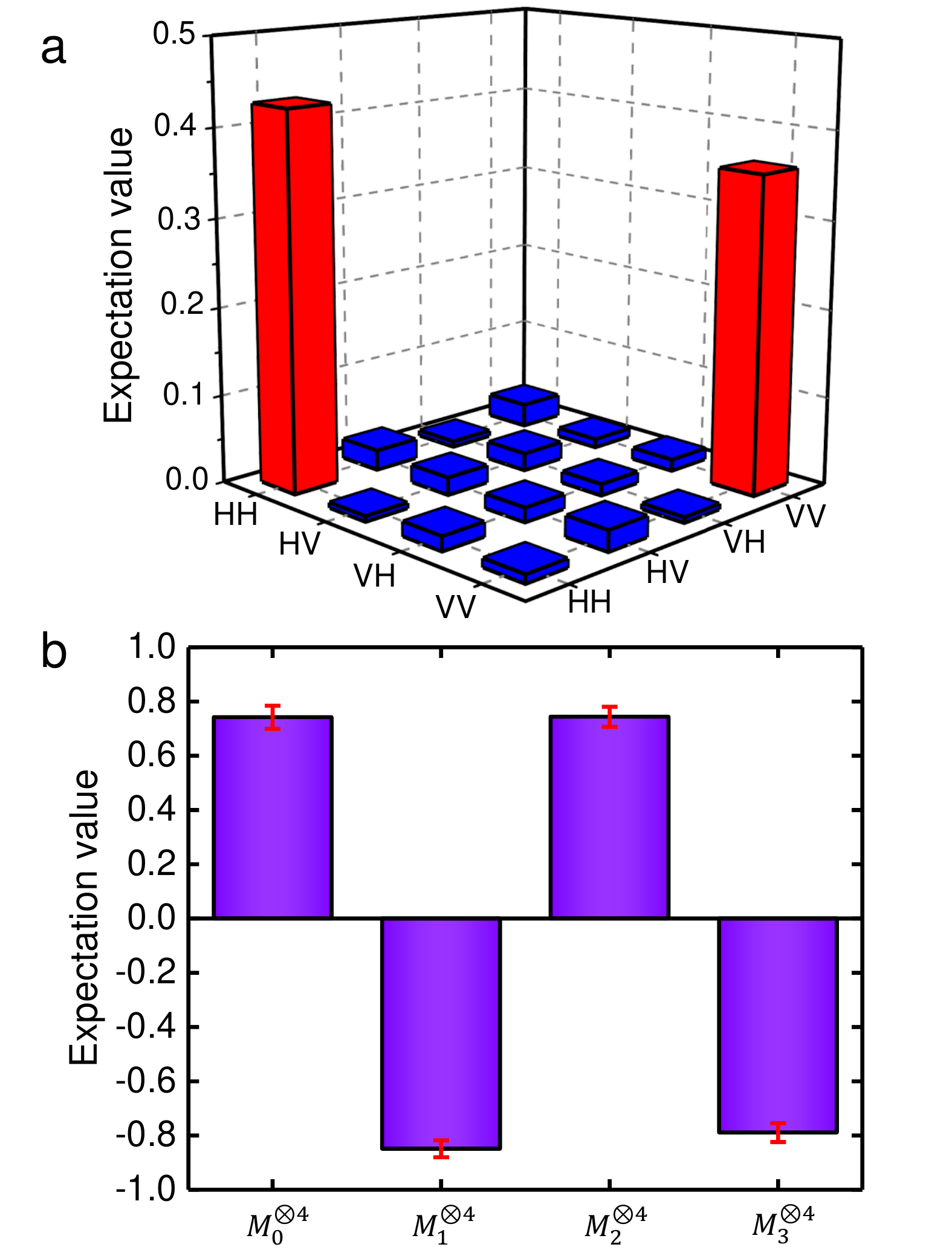}\\
  \caption{\textbf{Fidelity of the four-qubit GHZ state.}
  \textbf{a}, Normalized four-photon coincidence probabilities measured in the $H/V$ basis for the first term of Eq.~(\ref{fidelity}).
  \textbf{b}, Measured average values for the observables $M_n^{\otimes 4}$ ($n=0,\,1,\,2,\,3$) for the last term of Eq.~(\ref{fidelity}). For each measurement setting we record four-photon coincidence counts for one hour. Error bars represent one standard deviation.
  }
\end{figure}

\emph{\textbf{Performance.}}
To characterize the quality of the created four-photon GHZ state, we measure the fidelity by decomposing the density operator of an ideal GHZ state into local observables \cite{PhysRevA.76.030305}
\begin{align}
  & |\mathrm{GHZ}_4\rangle \langle \mathrm{GHZ}_4 | = \frac{1}{2} \big(|HHHH\rangle\langle HHHH| \nonumber\\
  & \qquad +|VVVV\rangle\langle VVVV|\big)+\frac{1}{8}\sum_{n=0}^{3} (-1)^n M_n^{\otimes 4}, \label{fidelity}
\end{align}
where $M_n\equiv \cos(n\pi/4)\sigma_x+\sin(n\pi/4)\sigma_y$ ($n=0,\,1,\,2,\,3$) while $\sigma_x\equiv |H\rangle\langle V|+|V\rangle\langle H|$ and $\sigma_y\equiv -i|H\rangle\langle V|+i|V\rangle\langle H|$ are Pauli matrices in the $H/V$ basis. The fidelity of the created four-photon GHZ state is measured to be $78.3(1.5)\%$ when we set the excitation probabilities of both ensembles to $p=0.1\%$, as shown in Fig.~2. This result significantly surpasses the entanglement threshold of $50\%$ \cite{witness04prl,guhne2009entanglement}, and thus proves the existence of genuine four-partite entanglement.

Next we show that this scheme demonstrates a memory-enhanced scaling in the preparation efficiency of GHZ states, as evidenced by Fig.~3. Without loss of generality, we set the excitation probabilities $p$ to be equal for both atomic ensembles. By measuring the four-photon coincidence count rate in the $|HHHH\rangle$ state, which is one half the generation rate of the four-photon GHZ state, we observe a linear scaling $O(p)$ as we vary $p$ from 0.1\% to 0.8\%. The experimental data (black diamonds) agree well with the theoretical results with all the retrieval efficiencies, detection efficiencies and postselection operations included (blue curve) \cite{supplementary}. In contrast, for a scheme without memory enhancement, the two pairs of atom-photon entangled states need to be created in the two atomic ensembles simultaneously at the joint probability of $p^2$. We plot this scaling as the orange curve in Fig.~3 under the same retrieval efficiencies and imperfections for comparison. This improvement in the four-photon case from a quadratic scaling versus $p$ to a linear one demonstrates the key advantage of the memory-enhanced protocol.

\begin{figure}
  \centering
  \includegraphics[width=7.5cm]{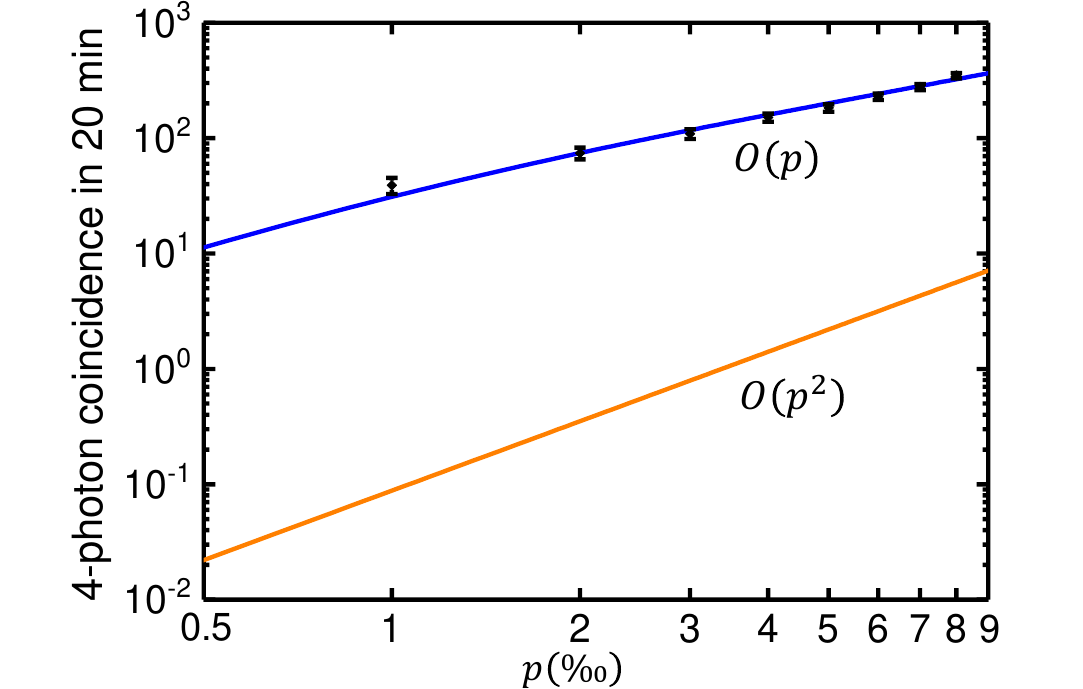}\\
  \caption{\textbf{Memory-enhanced scaling in generation efficiency of the four-qubit GHZ state.}
  Four-photon coincidence counts in 20 minutes are measured at various excitation probabilities $p$ ranging from $0.1\%$ to $0.8\%$ (black diamonds). Here we collect all the photons in the $H$ polarization, hence the coincidence rate gives us one half of the generation rate of the four-photon GHZ state. The blue solid curve is the theoretical four-photon coincidence rate in our experiment and the orange curve is the calculated coincidence rate for a protocol without memory enhancement \cite{supplementary}. These theoretical curves are calculated with all the retrieval efficiencies, detection efficiencies and postselection operations taken into account under the current experimental conditions. Error bars represent one standard deviation.}
\end{figure}

\begin{figure}
  \centering
  \includegraphics[width=7.5cm]{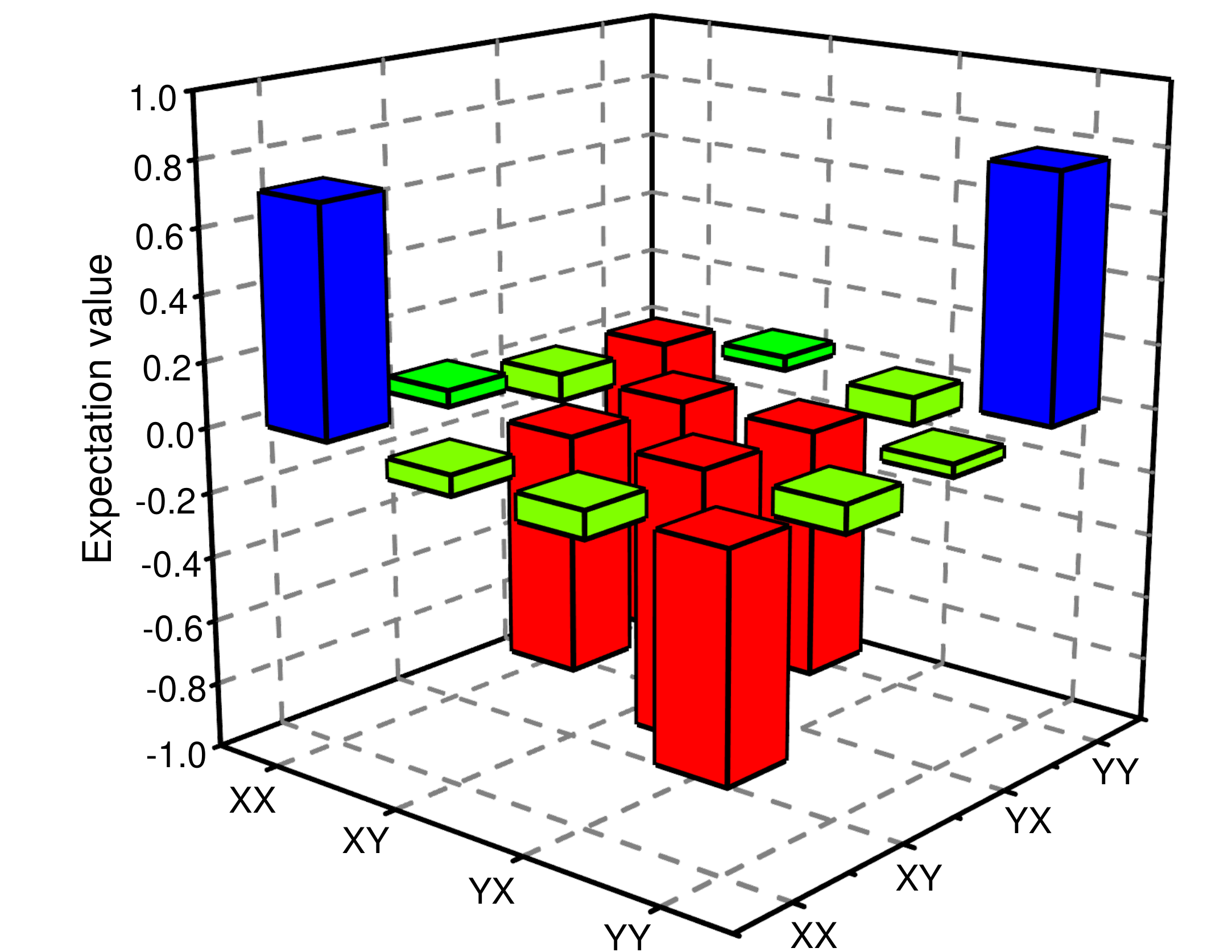}\\
  \caption{\textbf{Experimental test of the MABK inequalities and demonstration of quantum secret sharing.} We set the measurement basis of each qubit to be $\sigma_x$ or $\sigma_y$ and collect the four-photon coincidence counts under a fixed excitation rate for one hour each. The measurement basis is controlled by a QWP and an HWP to convert the $\sigma_x$ or $\sigma_y$ basis into the $\sigma_z$ basis of $H/V$ polarizations, followed by a PBS to separate the two polarizations into different paths to be detected by two SPDs. For each photonic qubit under a given measurement basis, we record $+1$ if it passes through the PBS and $-1$ if it is reflected. Here we plot the measured expectation values for all the 16 possible combinations of measurement bases (product of $\sigma_x/\sigma_y$ Pauli operators). For a given set of measurement bases, we record $M_+$ four-photon coincidence counts for the product to be $+1$ and $M_-$ coincidence for the product to be $-1$; then the expectation value is computed as $(M_+-M_-)/(M_++M_-)$. More details can be found in Supplementary Materials \cite{supplementary}.
   }
\end{figure}

\emph{\textbf{Applications.}}
We further demonstrate the applications of the prepared four-photon GHZ state in Bell-type inequalities and quantum cryptography. For $n=4$ qubits, a local realistic theory needs to satisfy the MABK inequalities \cite{MABK1,*MABK2,*MABK3}
\begin{align}\label{MABK}
&\langle F\rangle \equiv \left\langle \sigma_x \sigma_x \sigma_x \sigma_x \right\rangle + \left\langle \sigma_y \sigma_y \sigma_y \sigma_y \right\rangle \nonumber\\
&\quad - \left\langle \sigma_x \sigma_x \sigma_y \sigma_y \right\rangle
- \left\langle \sigma_x \sigma_y \sigma_x \sigma_y \right\rangle - \left\langle \sigma_x \sigma_y \sigma_y \sigma_x \right\rangle \nonumber\\
&\quad - \left\langle \sigma_y \sigma_x \sigma_x \sigma_y \right\rangle - \left\langle \sigma_y \sigma_x \sigma_y \sigma_x \right\rangle - \left\langle \sigma_y \sigma_y \sigma_x \sigma_x \right\rangle \le 2\sqrt{2}.
\end{align}
On the other hand, quantum mechanics allows a violation of the MABK inequalities with the largest possible violation given by the four-qubit GHZ state as $\langle \mathrm{GHZ}_4|F|\mathrm{GHZ}_4\rangle = 8$. To measure the expectation value of $F$, we evaluate each term in the expansion by choosing suitable $\sigma_x/\sigma_y$ measurement bases for each photon, as shown in Fig.~4. According to the experimental data and statistics, we estimate $\langle F\rangle = 6.01(0.11)$, which violates Eq.~(\ref{MABK}) by more than 28 times the standard deviation. Note that due to the postselection in our scheme, which is essential for the memory-enhanced scaling, the four photons of the GHZ state do not physically exist simultaneously. This will lead to the locality loophole \cite{RevModPhys.86.419} if we want to distinguish the quantum theory from local realism. Nevertheless, the violation of the MABK inequality can still verify the existence of quantum entanglement \cite{guhne2009entanglement}. Besides, if we accept the validity of quantum mechanics and just focus on the practical applications of the multipartite graph states such as measurement-based quantum computing \cite{oneway}, quantum secret sharing \cite{qss} and quantum networks \cite{kimble2008quantum,wehner2018quantum}, then the state prepared by our scheme produces exactly the same outcome as a conventional graph state whose qubits all exist at the same time. Below we consider one such application of four-partite quantum secret sharing \cite{qss} using the four-photon GHZ state.

Suppose Alice wants to send a message to Bob, Charlie and Dave in such a way that any two people cannot recover the message but all three together can. Equipped with four-photon GHZ states among the four people, a simple strategy is to measure their photons randomly in the $\sigma_x$ or $\sigma_y$ basis and publicly announce their choices. As evident from Fig.~4, when all their bases coincide, the product of the four outcomes will be +1 with high probability (for ideal GHZ states this probability is one). Similarly, when two measurement bases are $\sigma_x$ and the other two are $\sigma_y$, the product of the outcomes will be -1. The other choices of measurement bases will lead to no correlation and hence will be discarded. In this way, Bob, Charlie and Dave can jointly establish a shared secret key with Alice that can be used for encoding and decoding the message; but since each one's measurement outcome in the $\sigma_x$ or $\sigma_y$ basis is completely random, any two of the three people will not be able to get back the message. As shown in Fig.~4, we measure the four-photon coincidence in the 16 bases, pick out the 8 ones to be used for secret sharing and compute the quantum bit error rate (QBER) \cite{rmp02cryptography} which is the probability of deducing a wrong bit for the secret key due to the imperfect GHZ states. The QBER is estimated to be $12.46(0.66)\%$ (see Supplementary Materials), which is below the threshold of $15\%$ for the security against individual attack and allows classical error correction and privacy amplification to further improve the secret key \cite{rmp02cryptography}. Note that this QBER is still above the security threshold of $11\%$ for the coherent attack \cite{rmp02cryptography}, but it can still find practical applications assuming limited capacity of the eavesdropper \cite{rmp02cryptography,prl05qss}.


To sum up, we have demonstrated memory-enhanced preparation of four-qubit GHZ states with efficient scaling and have applied it for the violation of MABK inequalities and for quantum secret sharing. This change from a quadratic scaling to a linear one for preparing graph states, when generalized to larger number of qubits, leads to the substantial difference between an exponential scaling and a polynomial one \cite{supplementary}, thus opens up realistic prospects towards the future generation of large-scale multipartite entangled states in distributed systems with various applications in quantum information science and quantum metrology.

\begin{acknowledgments}
This work is supported by the National Key Research and Development Program of China (2020YFA0309500), the Frontier Science Center for Quantum Information of the Ministry of Education of China, and Tsinghua University Initiative Scientific Research Program. Y.K.W. acknowledges in addition support from Shuimu Tsinghua Scholar Program, International Postdoctoral Exchange Fellowship Program, and the start-up fund from Tsinghua University.
\end{acknowledgments}

%

\end{document}


\title{Supplementary Materials for ``Quantum-memory-enhanced preparation of nonlocal graph states''}
\author{Sheng Zhang}\thanks{These authors contributed equally to this work.}
\author{Yu-Kai Wu}\thanks{These authors contributed equally to this work.}
\author{Chang Li} \altaffiliation{Present address:  ISIS (UMR 7006), University of Strasbourg and CNRS, 67000 Strasbourg, France}
\author{Nan Jiang}\altaffiliation{Present address: Department of Physics, Beijing Normal University, Beijing 100875, China}
\author{Yun-Fei Pu}
\author{Lu-Ming Duan}\email{lmduan@tsinghua.edu.cn}
\affiliation{Center for Quantum Information, IIIS, Tsinghua University, Beijing 100084, PR China}
\maketitle

\subsection{Section S1. Long-lived atomic quantum memory}

To extend the lifetime of atomic quantum memories to the millisecond level, we start by cooling a sample of $^{87}$Rb atoms in a magneto-optical trap (MOT) and loading the atoms into a one-dimensional optical lattice. Then we pump the atoms into the upper clock state $|g\rangle=|5S_{1/2},F=2,m=0\rangle$ and use the magnetically insensitive hyperfine ground states $|g\rangle$ and $|s\rangle=|5S_{1/2},F=1,m=0\rangle$  for storage. Lifetime of tens of milliseconds is observed by the electromagnetically induced transparency (EIT) storage.

For a DLCZ-type atomic quantum memory, the retrieval efficiency and the cross-correlation function are two important parameters to characterize its quality. In our experiment, we observe that the retrieval efficiency drops from 8$\%$ to $4\%$ within the $1\,$ms storage time in QM1. The fast decay of the retrieval efficiency within $1\,$ms is due to the motion of atoms confined in individual optical lattice pancakes. For QM2, long-time storage is not used in our scheme so we only measure the retrieval efficiency as 4$\%$ when the storage time is 10$\,\mu$s. We also measure the cross-correlation $g_c$ and the anticorrelation parameter $\alpha$ for QM1 within the $1\,$ms storage time, as shown in Fig.~S1.

\begin{figure}[htbp]
  \centering
  \includegraphics[width=12cm]{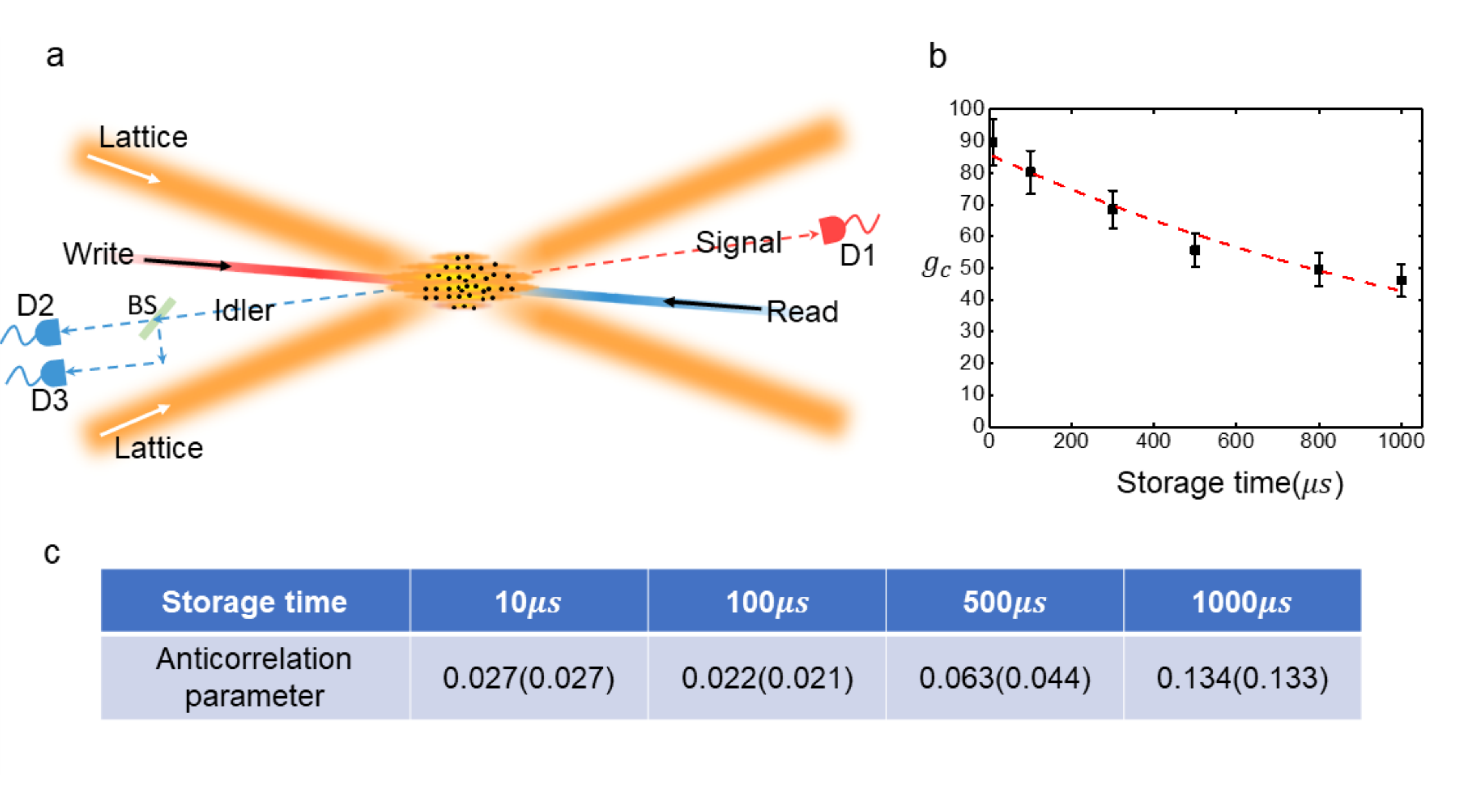}\\
  \caption{\textbf{Experimental setup for characterizing the long-lived atomic quantum memory QM1 and its performance.} \textbf{a}, $10^6$ atoms are loaded into a one-dimensional optical lattice. The write laser beam propagates in the opposite direction to the read laser beam. The signal photon is collected at a small angle of $1.5^\circ$ relative to the write laser beam. The idler photon is collected along the opposite direction to the signal photon. A beamsplitter (BS) is inserted into the path of the idler photon followed by two single photon detectors D2 and D3 to measure the photon anticorrelation.
  \textbf{b}, Decay of the cross-correlation function $g_c$ in QM1 within $1\,$ms. The cross-correlation function is defined as $g_c\equiv (p_{12}+p_{13})/[p_1(p_2+p_3)]$, where $p_i$ $(i=1,\,2,\,3)$ is the photon detection probability of the single photon detector $D_i$, and $p_{ij}$ is the signal-idler coincidence detection probability of single photon detectors $D_i$ and $D_j$. The measured cross-correlation function is far beyond the classical threshold of two after $1\,$ms storage, which implies that the nonclassical correlation between the signal and the idler photons is well preserved during the storage. The $1/e$ decay time is estimated to be $1.44(0.16)\,$ms. The presented data is for QM1 with an excitation probability $p=0.1\%$.
  \textbf{c,} The measured anticorrelation parameter for various storage time. The anticorrelation parameter is defined as $\alpha=p_1 p_{123}/(p_{12}p_{13})$, where $p_{123}$ is the three-fold coincidence count probability of detectors $D_1$, $D_2$ and $D_3$.  The presented data is for QM1 with an excitation probability $p=0.1\%$.}
  \end{figure}

\subsection{Section S2. Experimental timing sequence}

We use two National Instruments 6733 boards to load two samples of atomic ensembles into two optical lattices at the same time. After that, the experimental timing sequence begins with a trigger signal from the 6733 boards and is executed on a home-made-field-programmable gate array (FPGA). As shown in Fig.~S2, the experimental sequence starts with continually applying write pulses (W1) on QM1. If there is no signal photon (S1) click after a write pulse, a clean pulse identical to the read pulse is applied (not shown in Fig.~S2 for simplicity). Each write-clean cycle takes 1$\,\mu$s. Upon a signal photon click (S1) after a write pulse, the experimental sequence for QM1 halts and we begin to repeat the same write-clean trials on QM2 until there is a signal photon click (S2). Note that if there is no signal photon click (S2) within 1000 trials, the sequence will terminate and restart from QM1. This threshold of 1000 trials is based on a tradeoff between nearly deterministic generation of entanglement in QM2 and the small decoherence in QM1. Once we successfully register the signal photon (S2) from QM2, two read pulses are simultaneously turned on to retrieve the stored spin wave excitations. A four-photon coincidence count is recorded if the two idler photons can be detected after interference. Whether or not the four-photon coincidence count occurs, the sequence enters a new round. After $50\,$ms we terminate the whole experimental sequence and start again from loading atoms into the MOT and the optical lattice.

\begin{figure}[tbp]
  \centering
  \includegraphics[width=14cm]{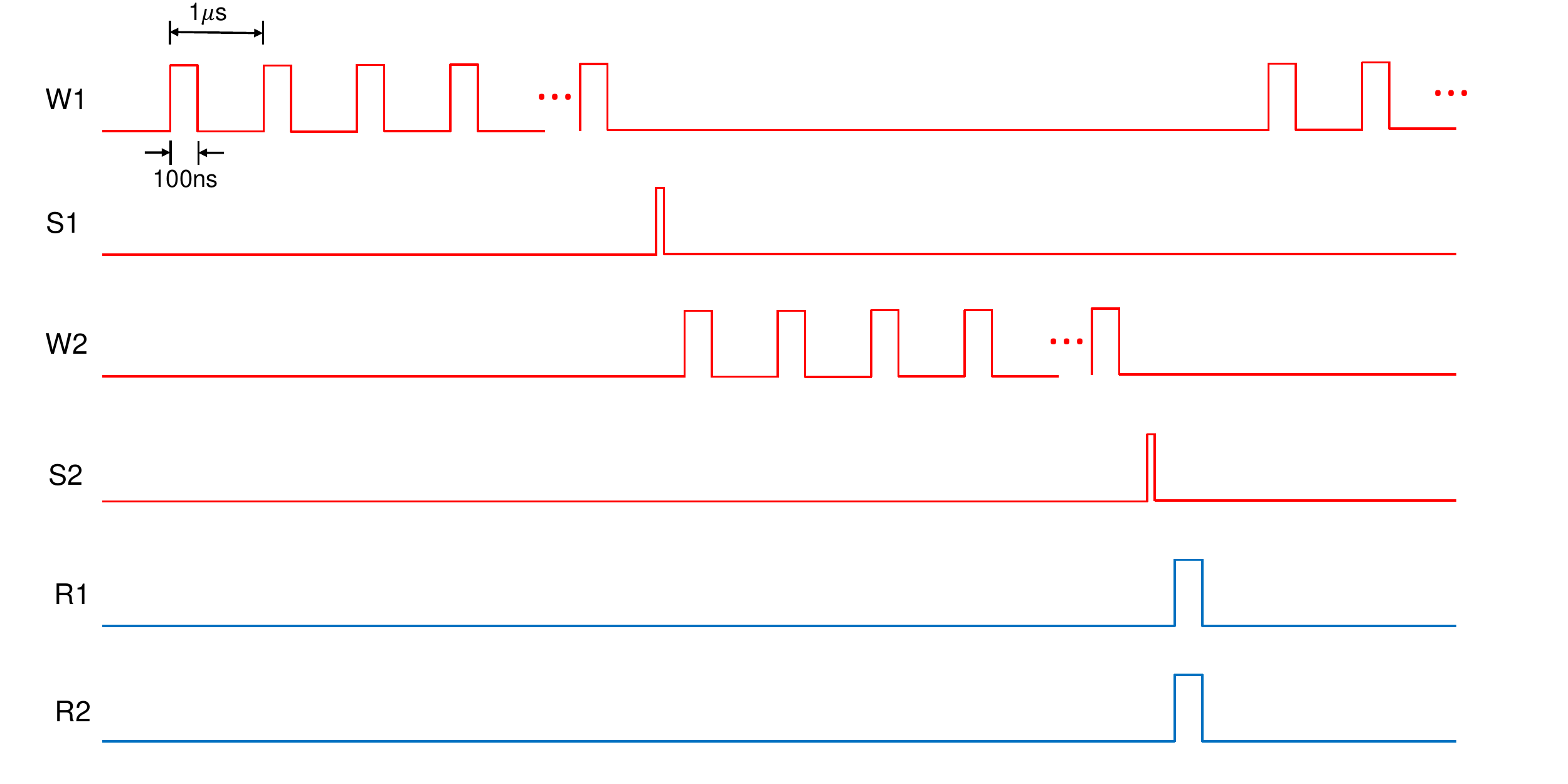}\\
  \caption{\textbf{Experimental sequence for the asynchronous preparation of two atom-photon entangled pairs.} }\label{}
\end{figure}

\subsection{Section S3. Entanglement fidelity of atom-photon entangled pairs}

To characterize the quality of each atom-photon entangled pair, we coherently convert the spin wave excitation  stored in the atomic quantum memory into a photon field and measure the entanglement fidelity by quantum state tomography. We reconstruct the density matrix of the signal-idler photon entangled state with maximum likelihood method, as shown in Fig.~S3a.  We further investigate the fidelity decay in QM1 versus storage time since it is used for long-time storage in our scheme. The fidelity drops from $96.2 (1.4)\%$ to $92.9(2.8)\%$ within $1\,$ms storage for QM1, as shown in Fig.~S3b. As QM2 is not involved in long-time entanglement storage, we just fix the storage time at 10$\,\mu$s and reconstruct the density matrix in Fig.~3a with a fidelity of $90.6(1.6)\%$.

\begin{figure}[htbp]
  \centering
  \includegraphics[width=16cm]{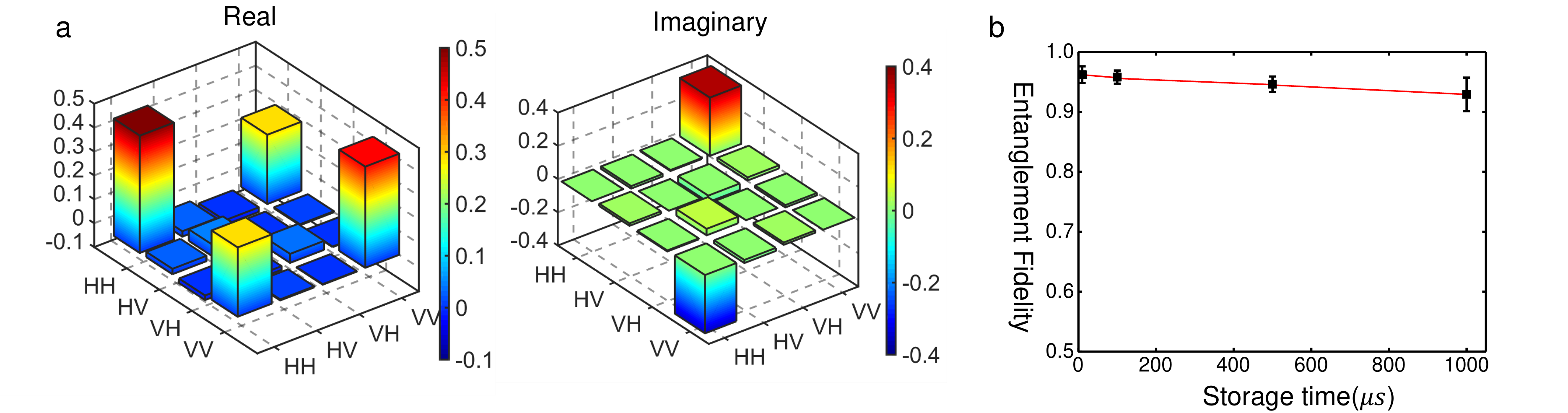}\\
  \caption{\textbf{Entanglement fidelity for atom-photon entangled pairs}.
  \textbf{a}, The reconstructed density matrix of the signal-idler photon entangled state in QM2 when the storage time is 10$\,\mu$s. The fidelity is estimated to be $90.6(1.6)\%$.
  \textbf{b}, Entanglement fidelity of the signal-idler entangled state in QM1 within $1\,$ms storage time.
  }\label{}
\end{figure}

\subsection{Section S4. Memory-enhanced scaling in generation efficiency of the four-qubit GHZ state}

To show the memory-enhanced scaling in the generation efficiency of the four-qubit GHZ states, we compare our protocol with a protocol without memory enhancement. For the protocol without memory enhancement, the two atom-photon entangled pairs need to be generated at the same time. Thus the joint success probability is proportional to $p^2$ if we assume that the excitation probability of individual atom-photon entangled pairs is $p$. In contrast, asynchronous preparation of two atom-photon entangled pairs with the help of long-lived atomic quantum memories will cost about $1/p+1/p=2/p$ trials, which is equivalent to a joint probability of two entangled atom-photon pairs of $p/2$. Apart from that, the two protocols connect the two entangled pairs into a four-qubit GHZ state in the same way. Therefore, there is a scaling change from $O(p^2)$ to $O(p)$ in the generation efficiency of four-qubit GHZ states in our protocol, as is observed experimentally in Fig.~3 of the main text.

Besides, it is also necessary to compare the four-qubit GHZ state fidelity of the two protocols. In our experiment, the infidelity of the four-qubit GHZ state mainly comes from the imperfect preparation of atom-photon entangled pairs and the imperfect photon interference, which are on the order of about $20\%$. This infidelity does not depend on whether we prepare the two atom-photon pairs simultaneously or asynchronously. The additional infidelity due to the storage in QM1 of less than $1\,$ms is only about $3.3\%$ according to Sec.~S3 and is much smaller than the dominant error sources. Therefore we expect that the fidelity of the four-qubit GHZ states should be comparable in the two protocols.

\subsection{Section S5. Quantum secret sharing}

We experimentally demonstrate a quantum secret sharing protocol using GHZ states in the main text, which is further explained in Fig.~S4a. We plot all the measured expectation values in different bases in Fig.~4 and here we further present the detailed values in Fig.~S4b. The results in the eight bases with even number of $\sigma_x$'s and $\sigma_y$'s are used for the quantum secret sharing task. The quantum bit error rate (QBER) is calculated as $R_{\mathrm{wrong}}/(R_{\mathrm{right}}+R_{\mathrm{wrong}})$, where $R_{\mathrm{right}}$ ($R_{\mathrm{wrong}}$) is the total number of correct (incorrect) key bits deduced under these eight bases.

\begin{figure}[htbp]
  \centering
  \includegraphics[width=16cm]{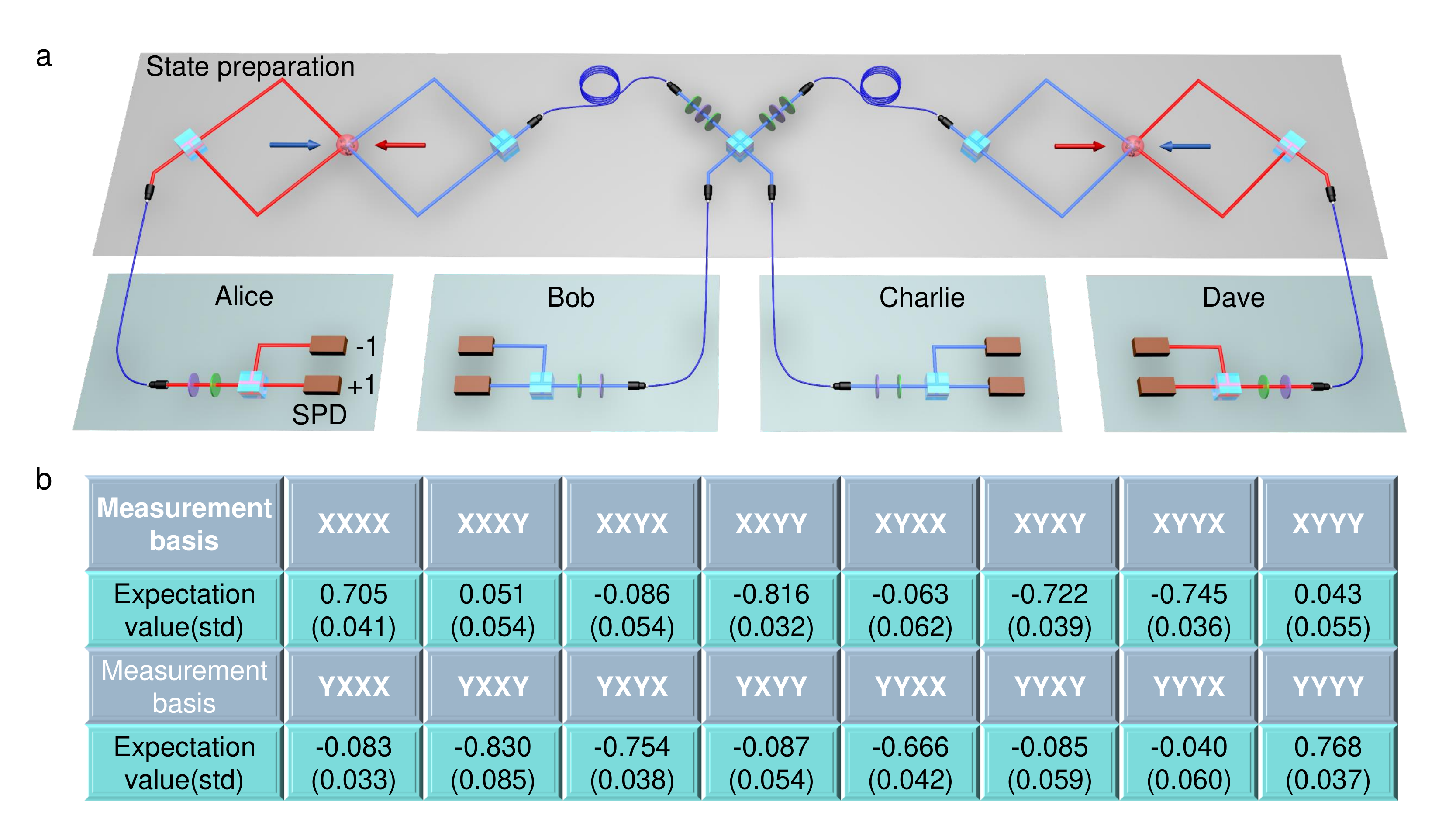}\\
  \caption{\textbf{Schematic diagram and experimental data for four-partite quantum secret sharing}.
  \textbf{a}, Each participant is assigned a photon from the four-photon GHZ state. A QWP, an HWP and a PBS constitute the measurement basis for each photonic qubit. For the randomly chosen $\sigma_x$ or $\sigma_y$ basis, each participant records +1 if the photon passes through the PBS and -1 if the photon is reflected.
  \textbf{b}, The measured expectation values in all the 16 possible combinations of $\sigma_x$ and $\sigma_y$ bases. Std represents one standard deviation.
  }\label{}
\end{figure}

\subsection{Section S6. Generalization to arbitrary graph states and scalability}
In this experiment, we demonstrate the simplest version of the scheme in Ref.~\cite{duan06prl} and explicitly show the memory-enhanced preparation efficiency. Here we briefly summarize the results of Ref.~\cite{duan06prl} about how this scheme can generate larger graph states and how the preparation time scales with the system size.

Suppose we have two independent graph states G1 and G2, each with $n$ qubits (in our experiment, G1 is the entangled state between photons 1 and 2, and G2 is that between photons 3 and 4). By applying a PBS gate, which consists of a polarization beam splitter (PBS3 in Fig.~1 of the main text) and a half-wave plate (which changes the local basis of individual qubits to meet the definition of graph states, and is not needed in this experiment for GHZ states), on qubit $i$ from G1 (photon 2) and qubit $j$ from G2 (photon 3), the two graph states can be combined together into a $2n$-qubit graph state, where qubit $j$ is now only connected to qubit $i$, while qubit $i$ is connected to qubit $j$ and all their neighboring qubits in the original graphs G1 and G2. By repeating this procedure as shown in Fig.~S5a and Fig.~S5b, larger graph states can be obtained.

The memory enhancement comes from the observation that if there are no additional operations on some of the qubits, their measurements can be expedited before the operations on the other qubits. This allows us to discard the unsuccessful events through postselection and to restart the preparation as soon as possible, while the succeeded parts can be held in quantum memories to wait for further connections. Suppose at each stage, we measure all the other qubits apart from the one used for later connections. Then the expected time to generate an $n$-qubit graph state is found to be \cite{duan06prl}
\begin{equation}
T \approx t_0 (\eta_s \eta_d)^{-1} n^{[(\log_2 n - 1) / 2 + \log_2 (1/\eta_d - 1 / 2)]}, \label{eq:S_scaling}
\end{equation}
where $t_0$ is the time of each attempt to generate an elementary entangled pair, $\eta_s$ the source efficiency and $\eta_d$ the detection efficiency. Compared with the notation in the main text, we have $p=\eta_s\eta_d$ as the measured excitation probability. Note that to obtain this scaling, after each PBS gate, one photon is measured for postselection while the other one should be stored again for later connections. This can be achieved by e.g. EIT storage using atomic ensembles which is compatible with our setup \cite{PhysRevX.9.041033}. Nevertheless, for the 4-photon GHZ state demonstrated in this work, since there is only one connection step, such an additional storage is not needed. From Eq.~(\ref{eq:S_scaling}), we see that the total preparation time scales as $T\sim n^{\log n}$ which is subexponential to the size of the graph state. In particular, this scaling is not affected by the finite detection efficiency $\eta_d$ of the practical single-photon detectors. Also note that, under a finite excitation rate $\eta_s$, we have about $O(\eta_s^2)$ probability to create double excitations in the elementary entangled pair, which is a fundamental source of infidelity in the prepared graph state. This error accumulates linearly with the qubit number. Therefore, to maintain a high state fidelity, $\eta_s$ should decrease as $1/n$ with the increasing system size \cite{duan2001long}. This contributes to a factor of $n$ in the preparation time and is again only polynomial with the system size.

In the original derivation of Ref.~\cite{duan06prl}, only the source and the detection efficiencies are considered, but the results can be easily generalized to include a finite retrieval efficiency $\eta_r$ of the quantum memories
\begin{equation}
T \approx t_0 \frac{\eta_r}{\eta_s \eta_d} \left[\frac{1-\eta_r/2-\eta_r\eta_d/4}{\eta_r\eta_d(1-\eta_r/2)}\right]^{\log_2 n} \left(\frac{2}{\eta_r}\right)^{(\log_2 n) (\log_2 n - 1) / 2}.
\end{equation}
In the limit $\eta_r\to 1$, we recover the result of Eq.~(\ref{eq:S_scaling}). Again, here we have a leading-order term of $(2/\eta_r)^{(\log_2 n) (\log_2 n - 1) / 2} \sim n^{\log n}$ which is subexponential in the system size.

Finally, note that in the above analysis we focus on tree-graph states (Fig.~S5a and S5b). By having more than one qubits in each subgraph for connections, we can also obtain more complicated graph states as shown in Fig.~S5c and Fig.~S5d, although the detailed scaling will then depend on the structure of these graphs.
\begin{figure}
  \centering
  \includegraphics[width=15cm]{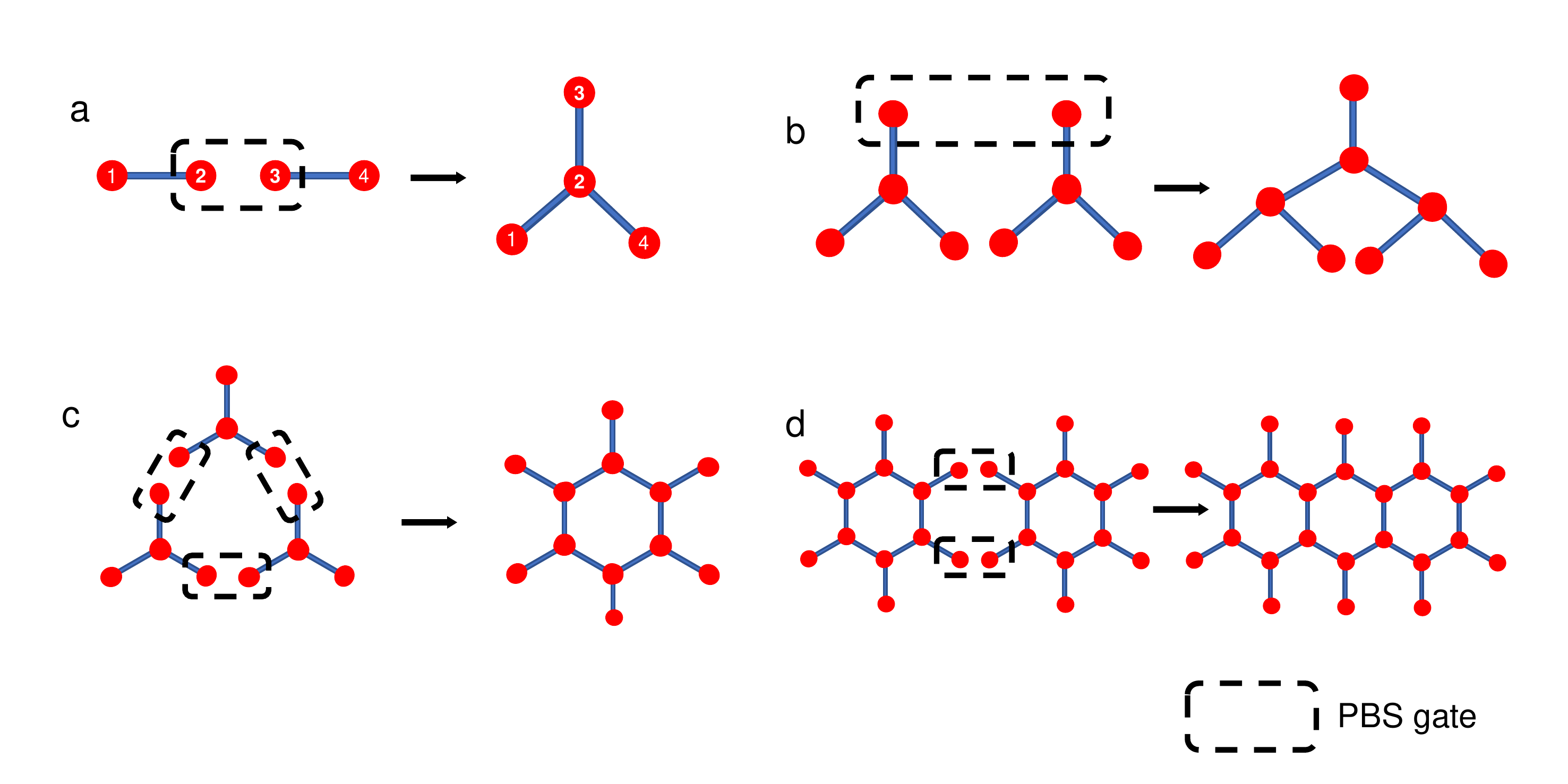}\\
  \caption{\textbf{Illustration of generating arbitrary graph states.} \textbf{a}, \textbf{b}, First two levels of generating tree-graph states. \textbf{c}, \textbf{d}, With more than one connection qubits in each subgraph, more complicated graph states can be obtained. \label{}}
\end{figure}

\subsection{Section S7. Calibration and compensation of relative phase in the GHZ state}
To compensate the relative phase $\phi$ in Eq.~(3) of the main text, we introduce another tunable phase shift $\phi^\prime$ between the $H$ component and the $V$ component by inserting a pair of quarter-wave plates parallel to each other at an angle of $45^\circ$ to the $H$/$V$ axes, and a half-wave plate in the middle. We measure the 4-photon coincidence counts under the correlation basis $(|H\rangle+|V\rangle)^{\otimes 4}/4$, which is proportional to $\cos(\phi-\phi^\prime)$. By fitting this curve as shown in Fig.~S6, we can thus set $\phi^\prime$ to compensate the relative phase $\phi$.
\begin{figure}
  \centering
  \includegraphics[width=10cm]{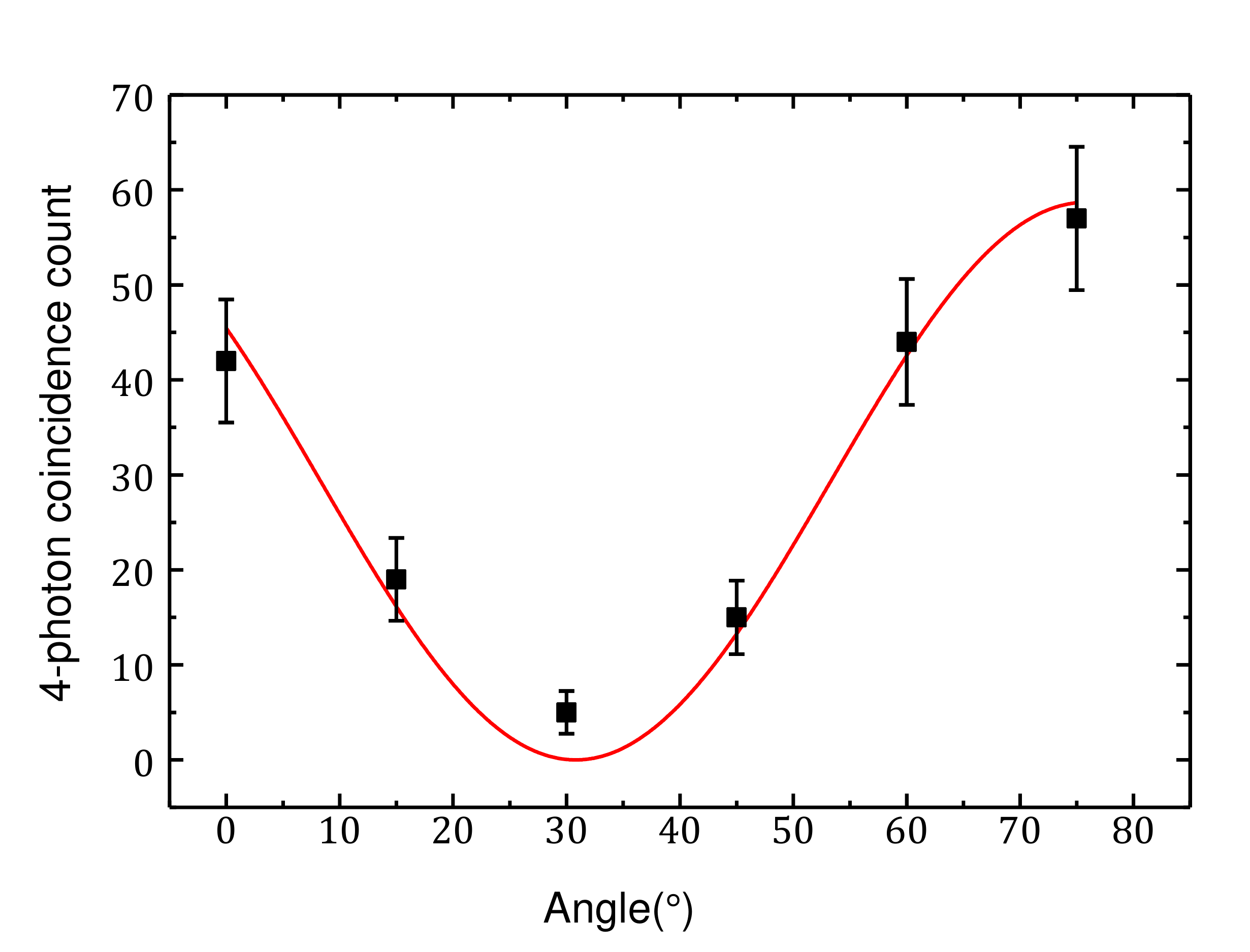}\\
  \caption{\textbf{Compensation of the relative phase $\phi$ in the GHZ state.} By rotating the angle $\theta$ of the half-wave plate, a tunable phase shift of $\phi^\prime=4\theta$ is achieved between the $H$ and the $V$ components, so that we observe a periodicity of $\pi/2$ in the experimental data. Fitted with a cosine function, we extract the maximizer of the curve at $\theta=75.8^\circ$, by which we can compensate the relative phase $\phi$ in the GHZ state.}\label{}
\end{figure}


%